\documentstyle[epsfig,12pt]{article}

\newcommand\PRL{Phys. Rev. Lett.}
\newcommand\PRC{{Phys. Rev.} C}

\newfam\BMath
\font\BMathL=cmmib10 
\font\BMathl=cmmib7
\font\BMathm=cmmib5
\textfont\BMath=\BMathL \scriptfont\BMath=\BMathl
\scriptscriptfont\BMath=\BMathm

\renewcommand\a{\alpha}

\renewcommand\d{\delta}

\newcommand\q{\theta}

\renewcommand\l{\lambda}

\newcommand\p{\pi}
\newcommand\r{\rho}

\renewcommand\o{\omega}

\newcommand\cs{{\cal S}}
\newcommand\ct{{\cal T}}   

\newcommand\ra{\rightarrow}

\newcommand{\half}{\frac{1}{2}}

\newcommand\be{\begin{equation}}
\newcommand\ee{\end{equation}}
\newcommand\bea{\begin{eqnarray}}
\newcommand\eea{\end{eqnarray}}

\newcommand\ba{\begin{array}}
\newcommand\ea{\end{array}}

\newcommand\eref[1]{Eq.~(\ref{#1})}

\newcommand\bfi{\begin{figure}}
\newcommand\efi{\end{figure}}
\newcommand\bpi[1]{\begin{picture}#1}
\newcommand\epi{\end{picture}}

\newcommand{\fref}[1]{Fig.~\ref{#1}}

\newcommand{\tref}[1]{Table~\ref{#1}}

\def\jou#1#2#3#4{{#1} {\bf #2} (#4) #3}

\def\qh{\q_{1/2}}

\font\lbigbf=cmbx12 scaled 1500

\begin{document}

\begin{flushright}
\small \sffamily NUC-MINN-00/06-T
\end{flushright}

\vspace{1.0cm}

\begin{center}
{\lbigbf Quantifying Baryon Stopping in High Energy}
\vspace{0.25cm}

{\lbigbf Nuclear Collisions}

\vspace{1.0cm}

{S.M.H. Wong}

\ \\
{\em School of Physics and Astronomy, University of Minnesota, Minneapolis, \\
Minnesota 55455, U.S.A.\footnote{email: swong@nucth1.hep.umn.edu}}

\end{center}

\vspace{1.5cm}
\begin{abstract}
We propose a numerical definition for baryon stopping in relativistic
heavy ion collisions that is obtainable from final hadron rapidity 
distributions as well as from bremsstrahlung measurements. 
Thus a new channel of communication is opened between the two methods.
\end{abstract}

\vspace{2.0cm}

\begin{flushleft}

\end{flushleft}

\vfill
\eject

\section{Introduction}
\label{s:intro}

Ever since the pioneering scattering experiments by Rutherford in the
early 1900s, there has been the persistent question of how matter would 
behave during very hard collisions. Two examples are the Landau picture 
where matter would come to a complete stop before exploding due to
the enormous pressure built-up or the Bjorken picture where target 
and projectile would pass through each other experiencing only partial 
deceleration in the process. Present day high energy nuclear collision
experiments provide not only an arena for settling this age-old 
question but also to see if there is any energy dependence that 
would change the collision scenario from one picture to another.
  
Another very good reason to study this is related to the original central 
goal of these experiments, which is to recreate deconfined matter, the 
so-called quark-gluon plasma, believed to have existed only in the 
early universe. To confirm its existence in the laboratory at the 
up-coming Relativistic Heavy Ion Collider (RHIC) at Brookhaven or the 
Large Hadron Collider (LHC) at CERN, one relies heavily on finding 
evidence amongst the many produced particles. These will be affected 
by the environment in which they are produced. In particular the net 
baryon number left in the central collision region or the amount of
baryon stopping affects, for example, photon and dilepton production. 
It is therefore of considerable interest to measure this in experiment. 
Indeed this has been done by NA49 at CERN's SPS \cite{na49} by measuring 
the net proton or baryon rapidity distribution of the final hadrons. 

Recently it was proposed \cite{jkcs} to measure this using the 
bremsstrahlung associated with slowing down of the baryons during the 
collisions. This method should be simpler than measuring hadron rapidities 
because of the extreme forward focus of the photons emitted from 
relativistically moving targets and projectiles. The measurement is 
therefore much more localized and only photons instead of different types 
of hadrons need to be detected. From a pragmatic point of view, 
it is better because most hadron detectors are not able to cover the 
full rapidity range. From a physical viewpoint, it can say something
about the space-time evolution of the collisions \cite{kstalk,kw}. This 
information will definitely not be available from rapidity measurements 
alone. In addition, it can distinguish between the Landau and Bjorken 
picture described above. 

In view of the two different ways of measuring baryon stopping, one would
like a way to compare the two and be able to communicate between them.
In any case, whenever one talks about baryon stopping today, one is usually
referring to the shape of the rapidity distribution. As far as we are
aware, there is no attempt to quantify it in any way so that the measurement
can be put into a more concrete footing. In this paper, we will make just
such an attempt. Whenever we refer to baryon stopping, we are referring
to stopping in rapidity space as is traditionally the case and not in
velocity space. It will be shown that stopping in rapidity and
velocity space are entirely different matters and therefore it is very 
important that it is made clear in which it is being described. There can 
be no confusion between the two.

\section{Quantifying Baryon Stopping or Transparency}
\label{s:st}

In order to quantify baryon stopping or its inverse, baryon transparency, 
we aim to find a quantity, called $\cs$, that satisfies the 
following requirements. 

\begin{itemize}

\item[(i)]{It should be equal to unity if there is complete stopping.
That is, all baryons end up having $y =0$.}

\item[(ii)]{If there is full transparency, and all final baryons move 
with the original initial rapidity $y_0$, this quantity should 
be zero.}

\item[(iii)]{One should be able to define its inverse, baryon transparency,
$\ct$ which is related to $\cs$ by the simple relation $\cs = 1-\ct$ so
that it has the opposite value in case (i) and (ii).}

\item[(iv)]{$\cs$ and $\ct$ should both be equal to half or approximately so  
when it is clear that there is half stopping and half transparency in 
rapidity space. For example, $\cs = \ct = 1/2$ when the rapidity distribution 
is totally flat.}

\item[(v)]{For different degrees of stopping, $\cs$ should range between
$0$ and $1$, signifying transparent to opaque in that order (or
the inverse for $\ct$).}

\item[(vi)]{It should be obtainable from both hadron rapidity data and from
bremsstrahlung measurements. This requirement is essential for bridging
the two types of measurements.} 

\end{itemize}

A quantity that satisfies all these is 
\be \cs = 1 - {{(1-v_0^2 \cos^2 \qh)} \over {2 v_0^2 \cos \qh}}
          \int^{+\infty}_{-\infty} dy {{v(y)\; \r(y)} \over {1-v(y) \cos \qh}}
\label{eq:sdef}
\ee
and therefore
\be \ct = {{(1-v_0^2 \cos^2 \qh)} \over {2 v_0^2 \cos \qh}}
          \int^{+\infty}_{-\infty} dy {{v(y)\; \r(y)} \over {1-v(y) \cos \qh}}  \; .
\label{eq:tdef}
\ee
Here $v_0= v(y_0)$ is the initial velocity in the center of mass 
frame which is related to the initial rapidity $y_0$ by the general
relation between velocity and rapidity 
\be  v(y) = \tanh y \; .
\ee
The $\r(y)$ is proportional to the final baryon rapidity distribution $dN/dy$. 
For symmetric target and projectile, it is defined by
\be \int^{+\infty}_{-\infty} dy \; \r(y) = 2.  
\label{eq:rdef}
\ee
To see that specifications (i), (ii) and (iii) are satisfied, one can consider 
two distributions
\be \r(y) = 2\; \d (y) 
\ee
and
\be \r(y) = \d(y-y_0) + \d(y+y_0) 
\ee
which correspond to full stopping and full transparency, respectively.
It is easily verified that one gets $\cs =1$, $\ct=0$ from the first 
distribution and $\cs =0$, $\ct=1$ from the second. 

Now for specification (iv), the flat rapidity distribution that respects
\eref{eq:rdef} would be
\bea \r(y) & = & 1/y_0  {\rm \hskip 0.9cm for }\; |y| \le y_0 
\label{eq:flatd}                                                           \\
           & = & 0      {\rm \hskip 1.5cm otherwise} \; .       \nonumber 
\eea
Using the formula
\be  \int {dy \over {1-\cos \q\; \tanh y}} = {y \over \sin^2 \q} 
     + {{\cos \q} \over {\sin^2 \q}} 
     \Big \{ \ln \cosh y + \ln (1-\cos \q\; \tanh y) \Big \} 
\ee 
and the requirement that $\cs = \ct = 1/2$ for a flat distribution, 
we end up with the equation
\be (2-v_0^2) \cos \qh - v_0^2 \cos^3 \qh                      
     - {{(1-v_0^2 \cos^2 \qh)} \over {y_0}} 
        \ln \Big ({{1+v_0 \cos \qh} \over {1-v_0 \cos \qh}} \Big ) 
     = 0  \; .
\ee
This equation defines the value of the angle $\qh$ which is the angle 
at which a flat rapidity distribution will yield $\cs = \ct = 1/2$
for a given $v_0$ or $y_0$. At this stage, $\qh$ is no more than a numerical 
quantity but its meaning will be explained below. In \tref{t:qh}, 
some values of $\qh$ have been computed at the various accelerators. The 
first case at SPS is for Pb+Pb and the second is for S+S collisions. The angle
becomes smaller as we go to higher energies. 

\begin{table}
\begin{center}
\begin{tabular}{||l|r|c|c|c||} \hline
Accelerator & $\sqrt{s}$/nucleon             & $1-v_0$  & $y_0$ & $\qh$  \\ 
            & [GeV]                          &          &       & (deg)\ \\ \hline
  SPS I     &   17.3 & 5.93$\times 10^{-3}$  & 2.91     & 28.83 \\
  SPS II    &   19.4 & 4.68$\times 10^{-3}$  & 3.03     & 26.97 \\
  RHIC      &  200.0 & 4.41$\times 10^{-5}$  & 5.36     &  7.93 \\
  LHC       & 1500.0 & 7.85$\times 10^{-7}$  & 7.38     &  2.87 \\ \hline
\end{tabular} 
\caption{The value of $\qh$ at the various accelerators.}
\label{t:qh}
\end{center}
\end{table}

\null
\bfi
\setlength{\unitlength}{1.5mm}
\begin{center}
\begin{picture}(60,20)
\put(5,5){\line(1,0){50}}
\put(30,2){\line(0,1){19}}
\put(7,5){\line(0,1){2}}
\put(53,5){\line(0,1){2}}
\put(3,2){$-y_0$}
\put(52,2){$y_0$}
\put(25,5){\line(0,1){13}}
\put(35,5){\line(0,1){13}}
\put(15,5){\line(0,1){13}}
\put(45,5){\line(0,1){13}}
\put(15,18){\line(1,0){10}}
\put(35,18){\line(1,0){10}}
\put(20,2){$-\l y_0$}
\put(33,2){$ \l y_0$}
\put(10,2){$-L y_0$}
\put(43,2){$ L y_0$}
\put(29,-1){$y$}
\put(26,23){$\r(y)$}
\put(-5,18){$\frac{1}{(L-\l)y_0}$}
\multiput(7,18)(2,0){5}{\line(1,0){1}}
\end{picture}
\end{center}
\caption{Simple test cases of rapidity distribution used in the text 
with $1 \ge L \ge \l \ge 0$.}
\label{f:test}
\efi

Our specification (iv) requires that rapidity distributions
intuitively half way between stopping and transparency in
rapidity space should be $0.5$. To show that this is indeed the 
case, we use a simple test distribution depicted in \fref{f:test}.
In this figure, we have two blocks at the height of $1/(L-\l)y_0$
because of \eref{eq:rdef} and symmetric about $y=0$. The parameters
$L$ and $\l$ allows us the freedom of a range of distributions. 
For our purpose here, we set $L=1-\l$ and vary the value of $\l$.
For any value of $\l$, we have two blocks symmetric about 
$\pm y_0/2$, and thus they should all have $\cs\sim\ct\sim 0.5$. 
We tabulated the value of $\cs$ for a set of values of $\l$ at the
various accelerators in \tref{t:half}. We recover the flat distribution
when $\l=0$ and we have two delta functions sitting at $\pm y_0/2$ 
when $\l=1/2$. We see that (vi) is better satisfied as we
go to higher and higher energies. Since this way of using bremsstrahlung 
will only be done at RHIC or at LHC, this is good enough
and we consider this specification met. In any case, they are all 
fairly close to $0.5$. 

One can introduce another special case which is without any doubt 
half way between opaque and transparent. That is
\be \r(y) = \half \Big ( \d(y-y_0) + \d(y+y_0) + 2\; \d(y_0) \Big ) \; .
\ee
This distribution is artificial but is ideal for our purpose here. This 
distribution describes half the baryons from the projectile and half
from the target sitting at $y=0$ and half of them from each initial
nucleus traveling with the original $y_0$. This is easily worked out to
give the exact result $\cs=\ct=1/2$. The purpose of this last distribution 
is to show that we have a sensible definition in hand.

\begin{table}
\begin{center}
\begin{tabular}{||c|c|c|c|c||} \hline
 $\l$     &  \multicolumn{4}{c||}{$\cs$}  \\ \cline{2-5}
          &   SPS I   &   SPS II  &    RHIC   &    LHC     \\ \hline  
 0        &  0.50000  &  0.50000  &  0.50000  &  0.50000   \\ 
 1/5      &  0.49552  &  0.49581  &  0.49921  &  0.49987   \\
 1/4      &  0.49354  &  0.49394  &  0.49881  &  0.49980   \\
 1/3      &  0.49028  &  0.49084  &  0.49801  &  0.49963   \\
 2/5      &  0.48820  &  0.48886  &  0.49836  &  0.49946   \\ 
 1/2      &  0.48689  &  0.48759  &  0.49689  &  0.49931   \\ \hline
\end{tabular} 
\caption{The value of $\cs$ for special test cases of $\r(y)$ 
depicted in \fref{f:test} with $L=1-\l$.}
\label{t:half}
\end{center}
\end{table}

\section{Difference Between Rapidity and Velocity Space}
\label{s:yvdiff}

When one looks for a numerical definition for stopping or transparency,
one encounters the question of whether this should be in rapidity or 
velocity space. If one was working within Newtonian mechanics, velocity
space would have been the automatic choice. This is quite logical since 
one could easily associate stopping with the slowing down of the incoming 
clusters of nucleons. However, it is also traditional to speak of baryon 
stopping while referring implicitly to the shape of $dN/dy$ in rapidity 
space. So it is in rapidity space that we gave this definition in the 
previous section. It must be stressed that the difference is huge between the 
two spaces. For example, the flat distribution given in \eref{eq:flatd}
at RHIC in rapidity space will appear as in \fref{f:fyv_vys} (a) in velocity 
space because
\be \r(y) = \frac{1}{y_0} \;\; \Longrightarrow \;\;
    \r(v) = \frac{1}{y_0 \;(1-v^2)}              \; .
\ee 
The halfway point of stopping in rapidity space at RHIC would appear 
to be much closer to transparent in velocity space! 
On the contrary, a flat distribution in velocity space 
\be \r(v) = \frac{1}{v_0} \;\; \Longrightarrow \;\;
    \r(y) = \frac{1}{v_0\;\cosh^2 y}
\ee
would be more opaque than transparent when one switches to rapidity. 
This is shown in \fref{f:fyv_vys} (b). Using this distribution, 
one finds $\cs \simeq 0.97$ or $\ct \simeq 0.03$.
\bfi
\centerline{\epsfig{figure=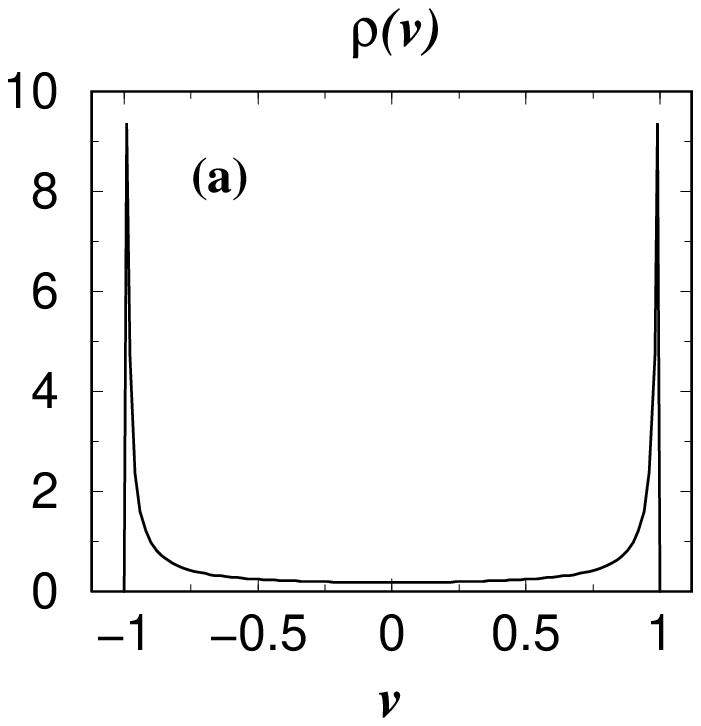,width=2.0in} \hskip 1.cm
            \epsfig{figure=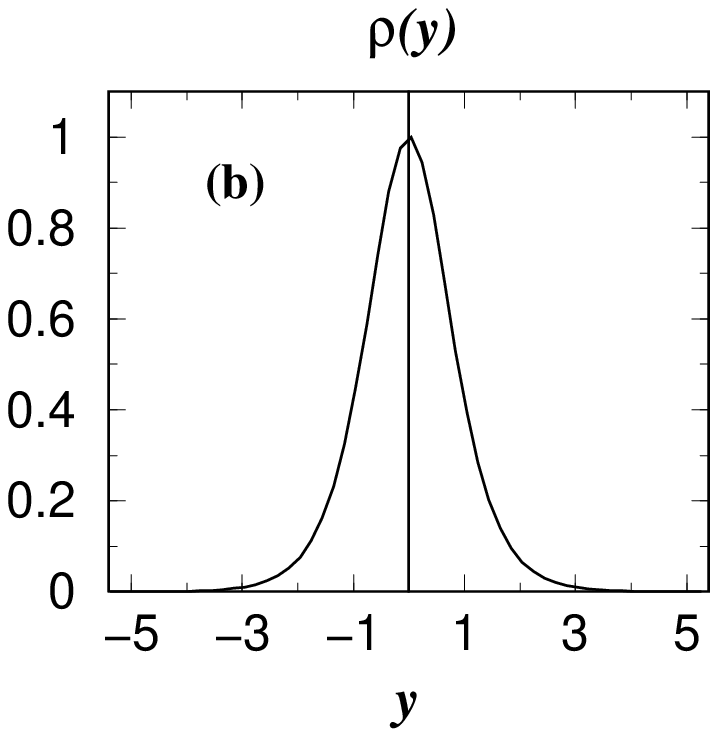,width=2.0in}}
\caption{(a) What a flat $\r(y)$ distribution looks like in velocity 
space and (b) a flat $\r(v)$ distribution looks opaque in 
rapidity space at RHIC.}
\label{f:fyv_vys}
\efi
\noindent For this reason, we have insisted that our definition be in 
rapidity space, conforming to the convention used in the heavy ion 
collision community.

\section{How To Obtain $\cs$ and $\ct$ From Bremsstrahlung Measurements}
\label{s:getcs}

In Sect. \ref{s:st}, the definition given for baryon stopping $\cs$ 
and transparency $\ct$ depended on the distribution $\r(y)$ which, 
for symmetric collisions, is related to the net baryon rapidity 
distribution by $dN_{B-\bar B}/dy = A\; \r(y)$ or net proton rapidity 
distribution by $dN_{p-\bar p}/dy \simeq Z\; \r(y)$. Here $Z$ and $A$ 
are the atomic and mass number respectively of the incoming target or 
projectile. This assumes that the net proton rapidity distribution is
representative of or approximately proportional to the net baryon 
distribution. This does not seem to be too bad an assumption if one
examines data from SPS \cite{na49}. Since one could easily calculate 
the value of $\cs$ and $\ct$ for some given data of $dN/dy$, 
the important question would be how to meet specification (vi) 
in Sect. \ref{s:st}. The main difficulty is how to obtain the quantity 
$\cs$ or $\ct$ from bremsstrahlung measurements. To solve this problem, we 
now state that $\cs$ can be related to the intensity distribution of the 
bremsstrahlung emitted in nuclear collisions by 
\be \cs = \frac{1-v_0^2 \cos^2 \qh}{v_0^2 \;\sin 2\,\qh} \;
    \Big ( \frac{4\p^2}{\a Z^2} \;
    \frac{d^2 I}{d\o\, d\Omega} \Big |_{{\o \ra 0\;\;\;\;} \atop {\q =\qh}} 
    \Big )^{1/2}      \; .
\label{eq:bd}
\ee
The numerical quantity $\qh$ has now been given the physical meaning of the 
opening angle from the beam pipe in which direction the soft photons should 
be measured. To see exactly where this angle lies in relation to other 
directions, one can work out the angle of maximum intensity assuming, for 
example, the case of full stopping $\;\cs=1$ or complete opacity $\ct=0$ 
using the formulae given in ref. \cite{kw}. In this case, the cosine of 
this angle $\q_{\rm{max}}$ is related to the initial velocity $v_0$ by 
\be \cos \q_{\rm{max}} = {(2-v_0^2)}^{-1/2}   \; . 
\ee
It then works out at RHIC to be $\q_{\rm{max}} = 0.538^{\,o}$ and
$\q_{\rm{max}} = 0.072^{\,o}$ at LHC. In view of the fact that the
intensity distribution falls off with the opening angle $\q$ from the
beam pipe, these $\q_{\rm{max}}$ are not too far from those $\qh$ in 
\tref{t:qh} at the respective accelerators so that there will be sufficient 
intensity at the $\qh$ to enable the photon measurements. 

If the reader has not guessed it already, we will now disclose the physical
meaning of the somewhat mysterious quantity $\cs$ or $\ct$. The inverse
of the prefactor to the rapidity integral in \eref{eq:tdef} is in fact 
proportional to the radiation amplitude for full stopping \cite{kw}. So 
one can now see why the rapidity integral itself will always be less than 
or equal to the inverse of this prefactor (but see the next paragraph 
concerning the soft photon requirement). Because of $\cs$ and $\ct$ have
an origin in the bremsstrahlung intensity distribution, they depend on 
the (charge) hadron rapidity distribution which automatically allows them 
to bridge the two different methods of determining baryon stopping. 
So far we have not mentioned the contribution from charged mesons to 
photon emissions. They could potentially ruin \eref{eq:bd}. However, 
pions are by far the most abundant meson type and they can be 
positively as well as negatively charged. Therefore contribution to
bremsstrahlung from mesons cancel out to a large extent \cite{us}.

Although it has been expressed in \eref{eq:bd} that the intensity 
distribution should be for low energy photons, in practice a few to  
tens of MeV should be good enough. The reason for the soft photon 
requirement is to remove all nuclear structural dependence as well as 
any potential interference effects. In ref. \cite{kw} it was shown 
that if there were more than one component in the acceleration of the 
nuclear clusters during the collisions, this would result in enhancement 
in and oscillations of the intensity distribution $dI/d\o\,d\Omega$ with 
$\o$. This is a direct result of the interference between the various 
components in the acceleration. For the purpose of our definition,
interference would unfortunately just taint any value of $\cs$ obtained
from \eref{eq:bd}. Only in the soft $\o$ limit is it free from this type 
of effect. As seen in Fig. 7 of ref. \cite{kw}, the intensity at small 
$\o$ is invariant under this.

\section{Examples}
\label{s:eg}

We will now try some example distributions and work out their $\cs$ and
$\ct$ values. For the test distribution in \fref{f:test} we vary the
two parameters $L$ and $\l$ to obtain the stopping values for the different
cases. The first five entries in \tref{t:eg} are for central single-block 
distributions centering around $y=0$. As the distribution is widened, $\cs$ 
decreases towards the flat $0.5$ value, as expected. The subsequent 
groups of entries are for symmetric two-block distributions shifting 
progressively away from the center to either side towards $|y|=1$. 
Thus in each group there is the tendency $\cs \ra 0$ and $\ct \ra 1$ 
which is how a sensible definition should behave. 
\begin{table}
\begin{center}
\begin{tabular}{||cccc|l|cccc||} \hline
 $L$      &    $\l$   &   $\cs$   &   $\ct$  &&  $L$      &    $\l$   &   $\cs$   &   $\ct$         \\ \hline 
 1/5      &     0     &   0.991   &   0.009  &&  1/2      &    1/3    &   0.700   &   0.300        \\ 
 1/3      &     0     &   0.965   &   0.035  &&  3/4      &    1/3    &   0.417   &   0.583        \\
 1/2      &     0     &   0.876   &   0.123  &&  4/5      &    1/3    &   0.377   &   0.622        \\ \cline{5-9}
 3/4      &     0     &   0.661   &   0.339  &&  3/4      &    1/2    &   0.229   &   0.771        \\ 
 4/5      &     0     &   0.622   &   0.377  &&  4/5      &    1/2    &   0.198   &   0.801        \\ \cline{1-4}
 1/3      &    1/5    &   0.926   &   0.074  &&   1       &    1/2    &   0.124   &   0.876        \\ \cline{5-9}
 1/2      &    1/5    &   0.800   &   0.200  &&  4/5      &    3/4    &   0.045   &   0.954        \\
 3/4      &    1/5    &   0.540   &   0.460  &&   1       &    3/4    &   0.018   &   0.982        \\ \cline{5-9} 
 4/5      &    1/5    &   0.499   &   0.501  &&   1       &    4/5    &   0.011   &   0.989        \\ \hline
\end{tabular}
\caption{The values of $\cs$ and $\ct$ for some more general test cases 
of $\r(y)$ depicted in \fref{f:test} with various $L$ and $\l$ at RHIC 
energies.}
\label{t:eg}
\end{center}
\end{table}

Admittedly these distributions are only test cases designed to show
how the numerical definition works. However, more realistic distributions 
can always be approximately reconstructed from thin blocks (strips) of 
varying heights so the simple distributions used do not affect in 
any way how the definition meets the specifications stipulated in 
Sec. \ref{s:st}. It may be that in practice our definition would need
to be refined but here we have laid the groundwork for a simple but
sensible numerical definition for baryon stopping. For the actual 
applications of this to real data, and for other ways of using 
bremsstrahlung from more realistic collision models than those used 
in \cite{kw}, we refer the reader to \cite{us}.

\section*{Acknowledgments}

The author would like to thank Joe Kapusta for a critical reading of
the manuscript and for discussions. This work was supported by the U.S. 
Department of Energy under grant DE-FG02-87ER40328.

\end{document}